\documentclass[modern]{aastex631}

\accepted{April 15, 2025}

\begin{document}

\title{Activity-Induced Near-Infrared Spectral Variability \\ at 29P/Schwassmann-Wachmann 1, 2017-2022}
\shorttitle{SW1 NIR Variability}
\shortauthors{Kareta et al.}


\author[0000-0003-1008-7499]{Theodore Kareta}
\affiliation{Lowell Observatory\\
1400 W. Mars Hill Road \\
Flagstaff, AZ, USA, 86001}
\author[0000-0003-1800-8521]{Charles A. Schambeau}
\affiliation{Florida Space Institute\\
University of Central Florida \\
Orlando, Florida, USA}
\affiliation{Department of Physics\\
University of Central Florida \\
Orlando, Florida, USA}
\author[0009-0007-6497-5156]{Megan Firgard}
\affiliation{Department of Physics\\
University of Central Florida \\
Orlando, Florida, USA}
\affiliation{Florida Space Institute\\
University of Central Florida \\
Orlando, Florida, USA}
\author[0000-0003-1156-9721]{Yanga R. Fern\'andez}
\affiliation{Department of Physics\\
University of Central Florida \\
Orlando, Florida, USA}
\affiliation{Florida Space Institute\\
University of Central Florida \\
Orlando, Florida, USA}

\begin{abstract}
29P/Schwassmann-Wachmann 1 (SW1) is both the first-discovered active Centaur and the most outburst-prone comet known. The nature of SW1's many outbursts, which regularly brighten the comet by five magnitudes or more, and what processes power them has been of particular interest since SW1's discovery in the 1920s. In this paper, we present and model four epochs of low-resolution near-infrared spectroscopy of SW1 taken with the NASA Infrared Telescope Facility and Lowell Discovery Telescope between 2017 and 2022. This dataset includes one large outburst, two periods of low activity (``quiescence" or ``quiescent activity"), and one mid-sized outburst a few days after one of the quiescent observations. The two quiescent epochs appear similar in both spectral slope and modeled grain size distributions, but the two outbursts are significantly different. We propose that the two can be reconciled if smaller dust grains are accelerated more than larger ones, such that observations closer to the onset of an outburst are more sensitive to the finer-grained dust on the outside of the expanding cloud of material. These outbursts can thus appear very rapid but there is still a period where the dust and gas are well-coupled. We find no strong evidence of water ice absorption features in any of our spectra, suggesting the areal abundance of ice-dominated grains is less than one percent. We conclude with a discussion of future modeling and monitoring efforts which might be able to further advance our understanding of this object's complicated activity patterns.

\end{abstract}


\section{Introduction} \label{sec:intro}
29P/Schwassmann-Wachmann 1 (SW1) is the first-discovered (but not the first recognized) Centaur, a class of Solar System small bodies on orbits crossing those of the Giant Planets while also not extending far beyond them. Some Centaurs display transient or periods of persistent comet-like mass loss, sometimes correlated with heliocentric distance and sometimes not, and some show modes of activity at unlike that seen at comets inside the water ice line whose mass loss is powered by water sublimation. Despite a discovery nearly a century ago in 1927, many basic questions about how SW1 behaves and how representative it might be of the Active Centaurs and Distant Comets in general remain unanswered or poorly constrained.

SW1 is best known for its massive outbursts (sudden changes in brightness brought on by changes in cometary activity) during which it can easily brighten in apparent visible magnitude by five magnitudes -- a factor of a hundred in flux -- in a matter of tens of hours or less (see, e.g., \citealt{1958PASP...70..272R, 1980AJ.....85..305W, 2008A&A...485..599T, 2016Icar..272..327M,schambeau_2017, schambeau_2019, 2020AJ....159..136W, morvan_2022}) several times a year. While some comets, like 17P/Holmes \citep{2008A&A...479L..45M}, and other Active Centaurs, like 174P/Echeclus \citep{2008PASP..120..393B}, show outbursts as large as SW1, none shows them as frequently. SW1's outbursts come in a range of strengths, from sub-magnitude sudden increases in brightness which dissipate within a day or two to multiple-magnitude events the debris from which can persist for weeks, and may even be quasi-periodic \citep{2008A&A...485..599T, 2010MNRAS.409.1682T}. Quiescent activity is the term often used to denote the lowest level of persistent activity seen on SW1, as the object has never appeared truly point-source-like and inactive despite a century of monitoring. Furthermore, SW1 might have both the dust-dominated outbursts seen at optical wavelengths as well as gas-only outbursts only detected through monitoring the object in the radio \citep{2020AJ....159..136W}, another kind of variation in outburst properties. While some of SW1's apparent uniqueness might stem from it being observed in more detail and over a longer timeframe than most active solar system small bodies due to its relatively bright apparent magnitude (V-mag typically $\lessapprox$ 17), some aspects of its activity behaviors do appear to be genuinely unique -- if there is another object in the Solar System that behaves like SW1, it has not been found.

SW1 does not appear to ever fully `shut off', stop shedding mass, and become point-source like in appearance (see, e.g., \citealt{1990ApJ...351..277J}). Even at its least active periods with a detected dust coma, CO$^+$ ions can be detected easily \citep{1980ApJ...238L..47L}, but these can also be detected during outbursts \citep{1980AJ.....85..474C} as well. While CO+ appears to be highly variable overall \citep{1982ApJ...254..816C}, the likely related ongoing release of carbon monoxide (CO) at SW1 \citep{1994Natur.371..229S, 2020AJ....159..136W, morvan_2022} appears to never fully shut off suggesting that the two are at least partially decoupled (see also \citealt{2019MNRAS.486.5614I}) likely due at least in part due to variable solar wind conditions \citep{1991Icar...90..172C}. The question of whether or not this is direct sublimation of carbon monoxide ice, release of carbon monoxide from some matrix of other material (e.g., amorphous water ice), conversion of (comparatively) less volatile carbon dioxide into carbon monoxide, or some other process entirely is a subject of significant ongoing theoretical work (see, e.g, \citealt{2022PSJ.....3..251L}). 

While the question of SW1's modern volatile reserves is still being investigated from multiple angles, SW1's ability to maintain a significant amount of them on long timescales is rather certain -- its nucleus is relatively large (effective spherical radius of $R_{NUC}=32.3\pm3.1$ kilometers; \cite{2021PSJ.....2..126S}) compared to typical Jupiter Family comets \citep{fernandez_2013, bauer_2017}. Another property that sets SW1 apart is its orbit. Compared to most comets, SW1's orbit is very low eccentricity ($e=0.044$ at time of writing\footnote{Unless stated otherwise, SW1's current orbital elements are taken from JPL Horizons orbit solution JPL K192/80 from 2023.}), resulting in comparatively small temperature swings from perihelion at $q=5.78$ AU to aphelion at $Q=6.32$ AU. Even when SW1 was in a higher eccentricity orbit ($e\approx0.13$) from the 1940s to the 1970s, it was still very outburst-prone \citep{1958PASP...70..272R}. SW1 is the prototypical object in the `Orbital Gateway' \citep{2019ApJ...883L..25S, 2020ApJ...904L..20S, 2023ApJ...942...92G}, a grouping of low-eccentricity orbits just exterior to Jupiter which could play a significant role in getting some objects on Centaur orbits into Jupiter Family Comet (JFC) orbits in the inner Solar System. Longer-term stays in Gateway-like orbits may induce significant differences in the near-surface layers of these objects which produce their activity. compared to a `typical' Centaur \citep{2023ApJ...942...92G}. It is often assumed that SW1's large size, rare orbit, and potentially long rotation period must play some role in explaining its explosive activity, but the role both properties play is not agreed upon.

It is thus of general interest to the comet and Active Centaur communities, as well as those who study the TNOs and JFCs, to understand what kind of mechanisms might power SW1's large outbursts, and how common or applicable these processes might be at other bodies. For the $\sim30\%$ of JFCs which pass through the Orbital Gateway \citep{2019ApJ...883L..25S} on their way to the Inner Solar System, they pass through the exact same thermal environment that is currently a part of what is powering SW1 to behave to violently. Given the challenges in directly detecting gas species outside of carbon monoxide or its dissociative byproducts, it is clear that other observational approaches must be attempted to make inroads in understanding its outbursts.

In this work we are primarily interested in what kinds of materials -- their composition, structure, and size distributions -- are ejected during SW1's large outbursts, how this material may differ from what is present around SW1 during periods of its quiescent activity, and how this information might help explain or constrain the mechanisms which power SW1's outburst-prone activity. In Section \ref{sec:obs}, we review our four separate epochs of SW1 spectra from the NASA Infrared Telescope Facility and Lowell Discovery Telescope. These spectra span two separate outbursts and two periods of quiescent activity, and we describe our assessment of how active SW1 was during each of our observations using archival ATLAS photometry. In Section \ref{sec:models}, we describe our modeling procedure and how well each of the kinds of models can be applied to each spectrum. In Section \ref{sec:discussion}, we tie together our modeling efforts and their observational circumstances to explore what scenarios might explain the spectral variation we see, and we end with Section \ref{sec:summary}, wherein summarize our observations and conclusions.

\section{Observations} \label{sec:obs}
\subsection{Observation Log and Activity Context} 
This work presents new high-quality near-infrared spectra of the inner coma of 29P/Schwassmann-Wachmann 1 taken throughout the 2017-2022 timeframe with the aim of understanding the spectral behavior of the coma during quiescent activity and during and after outbursts. Three of the four nights of observations were taken with \textit{SpeX} \citep{2003PASP..115..362R}, a low-to-medium resolution near-infrared spectrograph and imager on the NASA Infrared Telescope Facility (IRTF). Observations were taken in the low-resolution `prism' mode with a $0.8\arcsec$-wide slit, resulting in $R\sim200$ spectra over the $0.8-2.5\mu{m}$ wavelength range. One of the four nights of data was taken with the \textit{NIHTS} (Near-Infrared High-Throughput Spectrograph, \citealt{2021PASP..133c5001G}) on the Lowell Discovery Telescope (LDT). These observations were taken through the $1.3\arcsec$-wide slit, resulting in variable resolution ($R\sim200$ degrading to $R\sim80$ with increasing wavelength) over the $1.0-2.3\mu{m}$ range. The SpeX observations were obtained with the longest possible slit length of $60\arcsec$, while NIHTS has a fixed slit length of $12\arcsec$. The NIHTS observations were obtained during a period of relative quiescent activity (see the paragraph after next), so the coma was not too extended (visual coma extent of $\approx5-6\arcsec$, near-IR extent of $\approx1.5-2\arcsec$) and thus the smaller slit length is not expected to impact the comparisons of the SpeX vs. NIHTS spectra.. All observations from both instruments were reduced and extracted with the \textit{spextool} package \citep{2004PASP..116..362C}, designed originally for SpeX and then modified for use with NIHTS due to the similarities between the instruments. Observations of the comet were `book-ended' by observations of an angularly-close G-type star (less than 1 degree away) for telluric calibration and then finally slope-corrected through low-airmass observations of the well-studied Solar Analog SAO 93936 / Hyades 64. All observations were reduced and extracted with the same pipeline and then calibrated (e.g., first divided by the local standard for an airmass correction and multiplied by the ratio of the local standard to the Solar Analog to produce a comet divided by Solar Analog ratio spectrum) against the same Solar Analog to ensure a consistent set of observations that are uniformly calibrated.

\begin{figure}[ht!]
\plotone{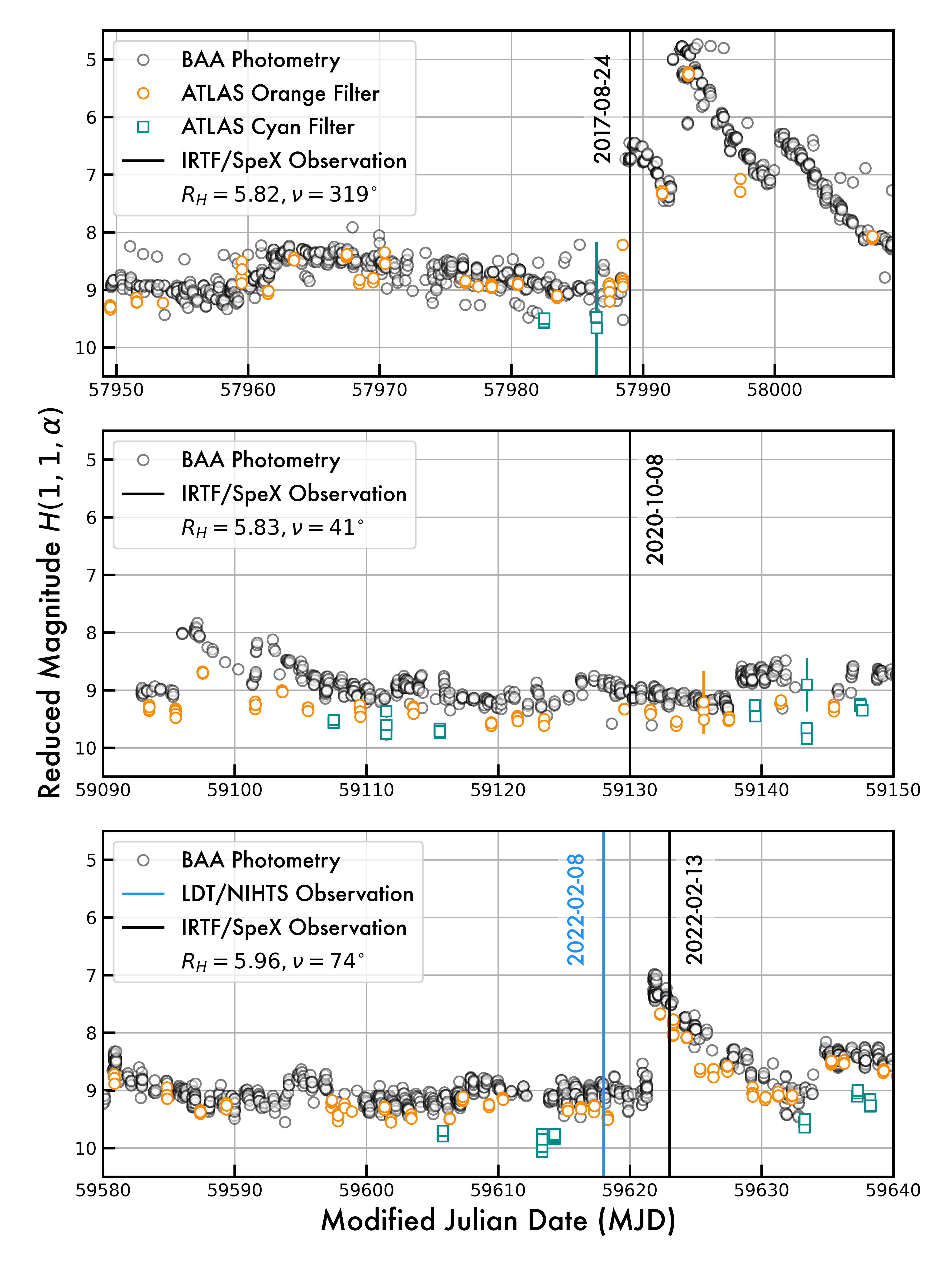}
\caption{Photometric measurements of 29P/Schwassmann-Wachmann 1 are shown from the ATLAS forced photometry server and the British Astronomical Association, corrected for heliocentric and geocentric distances but not for phase angle, relative to the dates on which we obtained near-infrared spectra of the comet. Observations with IRTF/SpeX are shown as black lines in each of the three epochs, and our LDT/NIHTS observations are shown as a light blue line in the bottom-most panel. The top panel corresponds to a large $\sim4$ magnitude outburst in 2017 August (actually two outbursts in rapid succession), the middle panel to a period of a minimum of brightness of the comet in 2020 October though, and the bottom shows a pair of observations before and after a moderate $\sim1.3$ magnitude outburst in 2022 February.}
\label{fig:log}
\end{figure}

SW1 is notoriously variable, so it is critical to place snapshot observations of the target in a more long-term context. In Figure \ref{fig:log}, we show when we obtained our spectra against photometry in two filters from the \textit{ATLAS} (Asteroid Terrestrial-impact Last Alert System, \citealt{2018PASP..130f4505T}) as obtained by their ``forced photometry server"\footnote{https://fallingstar-data.com/forcedphot/} \citep{2021TNSAN...7....1S}. While the full reduction and processing details for ATLAS are available in \citet{2018AJ....156..241H} and \citet{2020PASP..132h5002S}, we note that the forced photometry server does not use a fixed aperture size and instead optimizes their aperture size for best results for point sources. We also plot photometric observations obtained and submitted to the British Astronomical Association (BAA, \citealt{2022DPS....5441403M}) as reduced with a common $6$" radius circular aperture. These data come from a large number of observers with different telescopes and approaches, but are in quite good agreement as can be seen in Figure \ref{fig:log}. Given the relatively large pixel size ($1.86$") used by ATLAS, the typical $\sim2-4$ pixel radius used by the ATLAS forced photometry server is relatively comparable to the more numerous but heterogeneous observations by the BAA.

\startlongtable
\begin{deluxetable*}{c|c|c|c|c|c}
\tablecaption{Log of Spectroscopic Observations}
    \label{tab:log}
    \centering
    \startdata
        Date (YMD) & Telescope/Inst. & Seeing (Vis.) & N$_{exp}$, T$_{exp}$ & Airmass Range & Comments\\
        \hline
         2017 Aug. 24.40-24.44 & IRTF/SpeX & $\sim0.9$" & $28\times120s$ & 1.20$-$1.24 & Clear.\\
         2020 Oct. 08.43-08.65 & IRTF/SpeX & $\sim1.0$" & $28\times180s, 50\times200s$ & 1.01$-$1.50 & Clear.\\
         2022 Feb. 08.11-8.15 & LDT/NIHTS & $1.3-1.5$ & $16\times60s, 16\times90s$ & 1.01$-$1.04 & Clear.\\
         2022 Feb. 13.21-13.29 & IRTF/SpeX & $1.2$" & $26\times200s$ & 1.02$-$1.11 & Clear.
    \enddata
\end{deluxetable*}

The first epoch of observations happened on 2017 August 24. The \textit{SpeX} observations were obtained at the peak brightness of a $\sim2.5$ magnitude outburst which was followed by a second outburst of similar strength a few days later which together would result in the comet reaching an absolute magnitude of $H_V=5$ overall. The timing of this spectroscopic observation likely means that all of the dust released in this outburst was within (or very nearly so) the aperture used for our spectroscopic observations.

The second epoch of observations were obtained on 2020 October 8 during a period of quiescent activity. Inspection of spatial profiles obtained through stacking cometary and stellar images from the MORIS Guide Camera on the IRTF show that while the comet is certainly condensed in the inner $2-3\arcsec$ from the apparent nucleus, the comet can be clearly detected more than $\sim12\arcsec$ away, indicating the presence of an extended but faint dust coma. As can be seen in Figure \ref{fig:log}, the comet brightened slightly a few days prior to our observations but dimmed back to its quiescent minimum by the time our spectra were obtained. This same coma extension is not seen in the near-infrared spectra-derived spatial profiles, likely due to both the decreased spatial sensitivity of wavelength-dispersed spectra and from the decreasing sensitivity to detect small dust grains with wavelength.

The third and fourth epochs of observations were obtained on 2022 February 8 and 13 with LDT/NIHTS and IRTF/SpeX, respectively, sampling just before and a day after an outburst of SW1 of approximately $\sim1.3$ magnitudes. The pre-outburst observations were at a time of comparatively stable brightness and similar overall brightness to the 2020 October quiescent observations. Unlike the IRTF observations, SW1 was observed at similar enough airmass to the Solar Analog for the NIHTS observations that we just corrected the comet directly by the analog. The post-outburst observations came a day after peak brightness of the coma, suggesting that the coma had thinned some and thus optical depth effects might be less important. This is not to say that the 2017 and 2022 outbursts need to have ejected similar materials -- their relative strengths suggest that there should to be at least some differences -- this is only to say that our timing alone of when our observations were obtained relative to the peak brightnesses of the comet after both outbursts could induce some variation on its own.

\subsection{Comparing Spectra}
The four near-infrared spectra of SW1 taken during two outbursts and two periods of quiescent activity are shown in Figure \ref{fig:models} as black lines. The 2017 August spectrum taken early in an $\sim4$ magnitude outburst is the reddest and most concave of the reflectance spectra shown, being red and linear from $\sim0.8\mu{m}$ to at least $\sim1.6\mu{m}$ before becoming flat after $\sim2.0\mu{m}$. The second reddest spectrum is the October 2020 spectrum taken during quiescent activity, being both less red at short wavelengths and becoming neutrally sloped at $\sim 1.6\mu{m}$. The quiescent spectrum taken a year and a half later in 2022 February is nearly identical in slope, with the only difference being slightly less red below $\sim1.4\mu{m}$, though this is only barely statistically significant. A dip in reflectance near $\sim2.0\mu{m}$ is almost certainly not water ice absorption but instead an artifact of incomplete removal of the telluric $CO_2$ signal. Perhaps most interestingly, the least red spectrum is that from the second outburst in our sample from a few days later in 2022 February. The spectrum is barely red through $\sim1.8\mu{m}$, reaches a reflectance maximum somewhere in the $1.77-1.95\mu{m}$ telluric band, and then is neutral-blue at longer wavelengths. This spectrum also has a dip near $\sim1.3\mu{m}$ which is a telluric feature we could not correct fully and do not discuss further. The dip in reflectance in \textit{this} spectrum near $\sim2.0\mu{m}$ is not replicated in the other SpeX spectra, and thus might be real and thus possibly attributable to water ice. However, none of the spectra shows obvious, unambiguous signs of water ice -- they are all dust-dominated.

The similarity in overall slope between the two quiescent spectra taken at two different times with two different instruments suggests that SW1's reflective behavior during its quiescent activity is apparently stable, though significantly more observations of SW1 would be needed to understand just how stable it is. When SW1 is at its dimmest and presumably its least active, the kinds of dust that is in its immediate vicinity appears relatively similar across the time period studied. We leave a more significant discussion of this for Section \ref{sec:models} and \ref{sec:discussion}, where we attempt to quantify just how similar the two spectra are from a modeling perspective.

In contrast, the two outburst spectra appear very different, being the most red and most neutral in our dataset. While spectral curvature of the type seen in the 2017 August outburst spectrum (spectral slope decreasing with increasing wavelength) is seen at some level in all the spectra, it is most obvious in the two outburst spectra. The 2022 February outburst, while more similar to the quiescent spectra than the 2017 August outburst, even has a \textit{negative} spectral slope beyond $\approx2.0\mu{m}$, unlike any of the other spectra. The dust seen in the two outbursts clearly reflects light very differently, which is a significant focus of the rest of the paper.

We note that while our calibration scheme (same pipeline, same stars, similar instruments, same procedures) should minimize systematic effects causing variation between our spectra, very small differences in near-infrared spectra of solar system objects should still be looked at skeptically \citep{2020ApJS..247...73M}.

\section{Spectral Modeling}\label{sec:models}
Modeling the reflective behavior of the dust in cometary comae without clear absorption features requires making a large number of assumptions which significantly impact our ability to interpret the properties of the scattering material in an absolute sense (see a longer discussion of this topic in \citealt{firgard_and_kareta}). Are the dust and ice intimately mixed at the grain level, or are the dust and ice grains physically separated? Is the reflecting area of the coma dominated by fine grains in the Mie optics regime (our application is based on \citet{2018ApJ...862L..16P}, though we note that \citet{1982ApJ...254..816C} also applied Mie theory to a SW1 outburst for similar reasons as in this work) or would it be better to use a \citet{2012tres.book.....H} style model appropriate for grains larger than the studied wavelengths? Both sets of models might fit an individual dataset adequately, but additional information would be needed to discern which physical scenario is more plausible. One should not assume that just because one \textit{can} fit the data with a single grain size that there is only one size of grain in the \textit{real} coma -- some common sense is required.

Our approach is to pick a small number of classes of model to run that can at least adequately fit each of the spectra such that trends between the observations can be discerned. In addition to the Hapke-vs.-Mie choice mentioned previously, one also needs to select a set of optical constants to be used. Models of the solid portion of cometary comae in this wavelength range generally use two components, a spectrally featureless, red, and low-albedo material to fit the continuum (in this case, we make the conventional choice of amorphous carbon) and that of water ice. We employ the constants for amorphous carbon obtained by \citet{1983PhDT........18E} and \citet{1991ApJ...377..526R}, and the water ice constants of \citet{2008Icar..197..307M}. The two sets of amorphous carbon constants were produced under different laboratory conditions, differ in redness, and have different spectral curvature -- it is thus likely that the models might prefer one set over another (again, see a discussion of this in \citealt{firgard_and_kareta}) but it is unclear if these differences are physical. (One presumes the amorphous carbon present in cometary grains has some differences from any laboratory specimen!)

All comparison of models to data was done through the Markov Chain Monte Carlo (MCMC) implementation of \citet{2013PASP..125..306F}. All runs were completed with 64 walkers, $10^6$ iterations each, and the `burn-in' section of the chain to be discarded was estimated as three times the autocorrelation length of whichever parameter took the longest to burn in. We included a `fudge' parameter "$log(f)$" to test if our errors (as generated by spextool, \citealt{2004PASP..116..362C}) were accurate in magnitude. In essence, we allow the MCMC to try making the errors larger by some fractional amount. This additional parameter was found to be negligible, $\approx10^{-4}$ or smaller, indicating that our errors were plausible to within that margin.

\begin{figure}[ht!]
\includegraphics[width=0.6\textwidth]{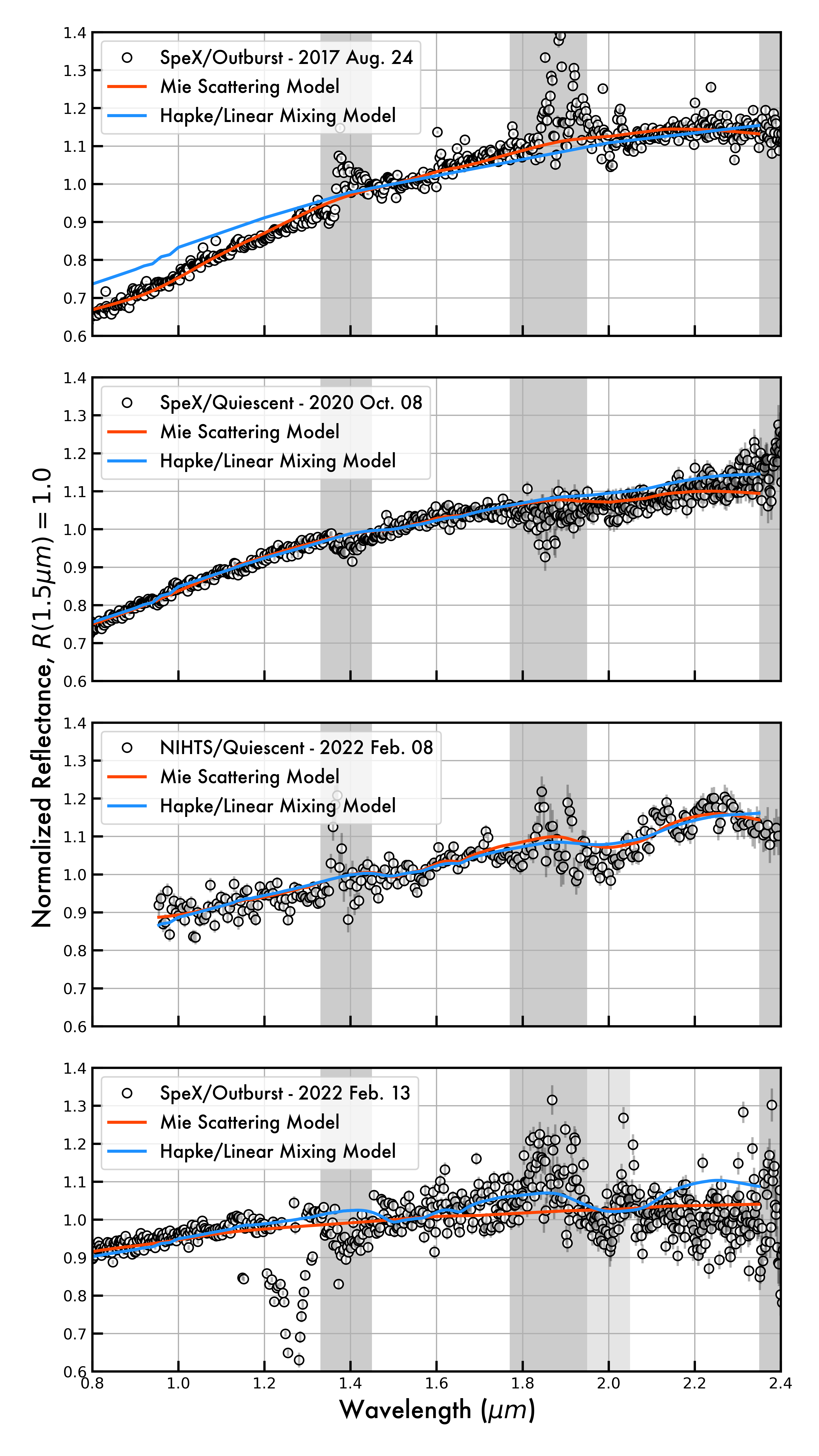}
\centering
\caption{The four low-resolution near-infrared spectra of Schwassmann-Wachmann 1 are shown as black unfilled circles compared against the two classes of models described throughout Section \ref{sec:models} shown as solid colored lines. As before, the light gray regions indicate areas of significant telluric absorption. The models for the 2022 February 8 spectrum are those derived from fitting the data excluding the uncorrected telluric feature near $\sim2.0\mu{m}$. The light blue models, labelled as `Hapke/Linear Mixing' models, assume large porous dust grains modeled by \citet{2012tres.book.....H} theory while the single-sized ice grains are pure and described by Mie theory. The orange-red models, labelled as `Mie Scattering' models, assume a power-law distribution of dust grains and a single size for ice grains, both described by Mie theory. More details on the modeling are available in the text. }
\label{fig:models}
\end{figure}

For both of the modeling methodologies described below, we fit all spectra from the two telescopes using an identical set of procedures. The only way in which this varied was over what wavelengths the model was compared to the data. The three spectra from the IRTF are of significantly higher SNR in wavelength regions of significant telluric absorption compared to the spectrum from the lower altitude LDT. For the IRTF spectra, we fit between $0.8-1.33\mu{m}$, $1.45-1.77\mu{m}$, and $1.95-2.35\mu{m}$ to avoid the areas of strongest telluric absorption. We refer to this as the `standard wavelength range.' We ran tests including the $1.33-1.45\mu{m}$ range for all of the four spectra, as that telluric band was often significantly better corrected than the $1.77-1.95\mu{m}$ band, especially in the two `quiescent' datasets. Our primary aim here was to use these different fitting ranges to assess the reliability of our datasets, and which conclusions were reliable regardless of how we sought to approach the data, similar in practice to our employment of two kinds of models. In summary, excluding both telluric bands almost always produced fits that converged quicker and had significantly lower $\chi^2$ values. For the LDT data, we tried this approach and also one that additionally excluded data in the $1.95-2.05\mu{m}$ range due to the challenges in correcting the telluric $CO_2$ absorption near those wavelengths, which we comment on below. The maximum likelihood models for each of the two classes of models are plotted against the four spectra in Figure \ref{fig:models} as solid colored lines. We first describe each model and its applicability to the data on its own and then compare the modeling approaches.

\subsection{Hapke-Style Large Grain Models}
\citet{2012tres.book.....H} models require that the grains of interest must be significantly larger than the wavelengths of scattered light studied. In our implementation of this model, we assume that only the dust grains are large enough for a Hapke approach and let the ice particles be of any single size (motivated by previous results suggesting an overabundance of ice near $D\sim1.0\mu{m}$, see, e.g., \citealt{2007Icar..190..284S,2018ApJ...862L..16P}). In order to allow any size of ice grains, we always model them with a Mie theory approach. This leaves us with four fitting parameters, the effective diameter ($D_{eff}$) of a typical dust grain, the porosity of that large dust grain (implemented as a fraction of the grain that is vacuum, $f_{vac}$, using \citealt{bruggeman1935berechnung} mixing), the diameter of an ice grain ($D_{ice}$) and the areal fraction of the reflecting coma that is taken up by these icy grains ($f_{ice}$). The allowed parameter values for these models are listed in Table \ref{tab:hapke_results} and are plotted in Figure \ref{fig:models} as light blue curves.

\startlongtable
\begin{deluxetable*}{c|c|c|c|c|c}
\tablecaption{Hapke-Style Model Results. All models were fit over the standard wavelength range except where noted, and parameter ranges reported are $16\%-84\%$ confidence intervals, analogous to 1-$\sigma$ errors.s}
    \label{tab:hapke_results}
    \centering
    \startdata
        Date (YMD) & $D_{eff}$ ($\mu$m) & $f_{vac}$ & $f_{ice}$ & $D_{ice}$ ($\mu$m) & Comments\\
        \hline
         2017-08-24 & $271_{-165}^{+159}$ & $0.04_{-0.02}^{+0.03}$ & $<0.001$ & $22_{-16}^{+19}$ & Very poor fit, $\chi^2=61.74$\\
         2020-10-08 & $234_{-172}^{+179}$ & $0.004_{-0.003}^{+0.007}$ & $<0.005$ & $8.0_{-0.4}^{+1}$ & $\chi^2=3.32$\\
         2022-02-08 & $238_{-171}^{+179}$ & $0.24_{-0.13}^{+0.11}$ & $0.007_{-0.002}^{+0.002}$ & $8_{-1}^{+5}$ & $\chi^2=2.99$\\
         ``  & $5_{-2}^{+308}$ & $0.44_{-0.18}^{+0.05}$ & $<0.005$ & $10_{-7}^{+23}$ & Excl. $1.95-2.05\mu{m}$, $\chi^2=2.79$\\
         2022-02-13 & $270_{-163}^{+156}$ & $0.49_{-0.01}^{+0.01}$ & $0.014_{-0.001}^{+0.002}$ & $1.7_{-0.4}^{+0.4}$ & $\chi^2=3.17$
    \enddata
\end{deluxetable*}

These models fit the quiescent spectra and the 2022 February outburst spectrum well but fit the 2017 outburst spectrum very poorly. Without clear absorption features to adapt to, Hapke models can in some situations become degenerate in grain size -- as has likely happened here. (As can be seen in Table \ref{tab:hapke_results}, the retrieved grain sizes for the dust and ice are often comparable their uncertainties for this reason.) The exception is that of our 2022 February outburst spectrum, which, while clearly unconstrained in dust size and at the maximum porosity that we allowed with our model, clearly prefers approximately $\sim1.4\%$ ice by area and an ice size of $D_{ice}=1.7\pm0.4\mu{m}$. The 2022 February quiescent spectrum has a maximum likelihood amount of ice of about $\sim0.7\%$, but this vanishes when the tellurically contaminated $1.95-2.05\mu{m}$ wavelength range is ignored. Within this class of models, the 2022 February outburst spectrum is the only one which shows evidence that water ice improves the quality of the spectral match that cannot be explained as an artifact easily. 

We also attempted to fit a simpler style of Hapke model to our datasets in which the dust and ice were intimately mixed into larger grains, with an effective diameter $D_{eff}$ and a ice fraction by weight $f_{ice}$. These models performed even worse on the 2017 August outburst spectrum, similarly or slightly worse on the two quiescent spectra, and worse on the 2022 February outburst. Notably, despite ice bands being absent from all but possibly the 2022 February outburst spectrum, several percent-by-mass of ice can be accommodated in both of the quiescent spectra without changing or perhaps even slightly improving the quality of model fits. The two quiescent spectra can handle $2-4\%$ ice by mass, while the 2022 February spectrum's maximum intimately-mixed ice content is harder to estimate. That model preferred values of $D_{eff}$ comparable to the wavelengths observed, which \citet{2012tres.book.....H} models do not formally allow, indicating that fines were still required at some level.

\subsection{Mie-Style Small Grain Models}
For this class of model, we assume that the amorphous carbon dust grains are physically separated from the icy grains as with the Hapke-style models, but the dust grains sizes are assumed to follow a power law distribution (with power law index $\alpha$ for the differential distribution) as a function of their radii. Grain radii between $0.5\mu{m}$ (e.g., about half that of our smallest wavelength) and $50\mu{m}$ are considered, and the range of allowed dust sizes was not varied in the fitting procedure.  Dust grain porosity of up to $50\%$ is implemented the same as before, which implicitly ignores any size-dependent porosity variation. This means that while for the Hapke-style models, we are fitting for the porosity of the large single-sized grains, here we are essentially fitting for a population-averaged porosity which may or may not reflect the properties of any given grain. The icy grains are assumed to be $100\%$ pure ice and single-sized, and all that is varied in the model (in addition to $f_{vac}$ and $\alpha$) is the areal fraction of ice in the coma ($f_{ice}$) and the size of the ice grains ($D_{ice}$). This specific model scheme is that of \citet{2018ApJ...862L..16P} with slight modifications described in \citet{2023PSJ.....4...85K}. All of these models used the amorphous carbon constants of \citet{1991ApJ...377..526R} as we found that, as in \citet{firgard_and_kareta}, the \citet{1983PhDT........18E} constants produced insufficiently red model spectra.

The results of our Mie scattering based models are reported in Table \ref{tab:mie_results} and are plotted in Figure \ref{fig:models} as orange-red curves. All of the spectra were well fit by this style of model and worse fit by Hapke models. The 2017 August outburst was best modeled by a steep size distribution of grains with $\alpha = 6.94_{-0.16}^{+0.21}$, the two quiescent spectra were intermediate in size distributions with $\alpha=5.03_{-0.07}^{+0.07}$ and $\alpha \sim 4-5$ in 2020 October and 2022 February respectively. While the model can fit the 2022 February outburst at wavelengths shorter than $\sim2.2\mu{m}$, the long-wavelength downturn seen in that spectrum cannot be reproduced by the model. As a result, that model moved to the lowest values of $\alpha$ of any of the models ($3.14_{-0.08}^{+0.11}$) and the highest porosity ($f_{vac}=0.469_{-0.044}^{+0.024}$), both near the edges of the allowed parameter search space.

\startlongtable
\begin{deluxetable*}{c|c|c|c|c|c}
\tablecaption{Mie-Style Model Results. All models were fit over the standard wavelength except where noted, and parameter ranges reported are $16\%-84\%$ confidence intervals, analogous to 1-$\sigma$ errors.}
    \label{tab:mie_results}
    \centering
    \startdata
        Date (YMD) & $\alpha$ & $f_{vac}$ & $f_{ice}$ & $D_{ice}$ ($\mu$m) & Comments\\
        \hline
         2017-08-24 & $6.94_{-0.16}^{+0.21}$ & $0.183_{-0.004}^{+0.004}$ & $<0.003$ & $21_{-14}^{+18}$ & $\chi^2=3.11$\\
         2020-10-08 & $5.03_{-0.07}^{+0.07}$ & $0.292_{-0.005}^{+0.005}$ & $0.005_{-0.001}^{+0.002}$ & $8_{-1}^{+6}$ & $\chi^2=1.95$\\
         2022-02-08 & $5.21_{-0.07}^{+3.25}$ & $0.115_{-0.06}^{+0.04}$ & $0.064_{-0.003}^{+0.053}$ & $1.7_{-0.6}^{+5.7}$ & $\chi^2=2.59$\\
         `` "" & $4.07_{-0.07}^{+0.20}$ & $0.118_{-0.02}^{+0.02}$ & $0.006_{-0.004}^{+0.014}$ & $8_{-6}^{+24}$ & Excl. $1.95-2.05\mu{m}$, $\chi^2=2.46$\\
         2022-02-13 & $3.14_{-0.08}^{+0.11}$ & $0.469_{-0.044}^{+0.024}$ & $<0.002$ & $18_{-14}^{+21}$ & $\chi^2=2.38$
    \enddata
\end{deluxetable*}

As mentioned previously, we tried fitting the NIHTS data with and without the likely telluric contamination between $1.95-2.05\mu{m}$, and this changed the $\alpha$ value somewhat and ice fraction $f_{ice}$ considerably. In general, inclusion of the suspect wavelengths resulted in models with non-neglible ice fractions being preferred (to fit the telluric $CO_2$ feature as solid water ice absorption), which then drove $\alpha$ to higher values to make the dust redder to counteract the abundance of bright blue pure ice. Considering that the models only find water ice in a significant way when the contaminated wavelengths are included, and that lower $\alpha$ models can fit the data just as well without ice, we think that the truth lies closer to the $\alpha\sim4.1$ and $f_{ice}\sim0.00$ case. However, the $2.2-2.35\mu{m}$ behavior of the NIHTS spectrum \textit{is} better fit by the inclusion of some ice, so like with the Hapke models described previously, some ice is compatible with the quiescent spectrum -- how much is not clear because of the telluric artifacts. 

Expanding beyond the comparatively narrow $16\%-84\%$ confidence interval reported in Table \ref{tab:mie_results}, the distributions of $\alpha$ values for the February 2022 quiescent spectrum and that from a year and a half earlier in 2020 October overlap considerably. We also note that if one includes the $1.33-1.45\mu{m}$ and $1.95-2.05\mu{m}$ regions for both quiescent spectra, the median $\alpha$ value for both is $4.86-4.87$. We thus conclude that while the two quiescent spectra have some slight differences in which combinations of parameters can model them best, the two are on the whole more similar than different, and thus that this model suggests that the comet was indeed behaving similarly at both epochs as might be inferred from the broadband photometry alone.

\subsection{Comparing Models}
The two classes of spectral model, with very different assumptions built in, tell a story with at least some commonalities. The 2017 August outburst spectrum, which the Mie scattering models found could be well described by a very steep size distribution of dust particles, was also the spectrum that the large-grained \citet{2012tres.book.....H} models performed the worst on. The 2022 February outburst spectrum was almost the opposite, where the Mie scattering models tried to reach the shallowest size distribution allowed by the fitting routine, was better fit by a large-grained Hapke model. In other words, both models indicate that the 2017 August outburst was fine-grain-dominated, and the opposite was true for the 2022 February outburst. Some of this might be due to using amorphous carbon as our opaque of choice, as other materials might have spectral behaviors which can reproduce both end-member spectra more easily. However, as comet dust really does have significant carbon content and previous studies (see, e.g., \citealt{2018ApJ...862L..16P}) have employed carbon in this way to good effect, we do not think this jeopardizes this conclusion.

The two quiescent spectra, similar in overall slope and taken when the comet was similarly dim, are reproduced by similar model parameters in both modeling frameworks. While the 2022 February quiescent spectrum is equally well modeled by either a Hapke or Mie framework, the 2020 October quiescent spectrum was slightly better modeled by a fine-grained Mie scattering model. Given that the 2022 February quiescent spectrum was of lower signal to noise ratio and had additional telluric contamination, the model results from the higher fidelity 2020 October data might be more reliable in this regard.

One last thing we note is that both classes of model had some challenges fitting the longest wavelengths in some of the spectra. While the long-wavelength decrease in reflectance beyond $\sim2.25\mu{m}$ in the NIHTS data could possibly be another imperfectly corrected telluric feature, the long wavelength behavior of the 2020 October and 2022 February SpeX datasets beyond $\sim2.00\mu{m}$ but shortward of $2.25\mu{m}$ is significantly more challenging to explain away. As all of the spectra were ratioed to the same solar analog, it seems that at least a significant amount of this variation is real -- and not accounted for by either modeling procedure. We view this as at least limited evidence for a more complicated distribution of grain sizes, compositions, and properties than was accounted for in either of our modeling scenarios. Investigating even more complex models is beyond the scope of such low-resolution spectra taken over such a limited wavelength range, but future higher SNR work over a longer wavelength range -- like that facilitated by JWST -- might be able to test some of these more complicated scenarios.

\section{Discussion}\label{sec:discussion}
\subsection{Outburst Properties}
The two outburst spectra obtained and modeled in this work cannot be reconciled as representing a similar physical state of the comet. The 2017 outburst spectrum, taken around the peak brightness of the first $\sim2.5$ magnitude outburst in a multiple outburst sequence, is our reddest spectrum and the 2022 outburst spectrum, taken after peak brightness of a smaller outburst, is our most neutral. The ATLAS and BAA data shown in Figure \ref{fig:log} also indicate that the quiescent periods prior to both outbursts were photometrically similar, and thus the pre-outburst coma environments around SW1 could have been rather similar as well.

The Hapke style models fit the 2017 August outburst rather poorly -- no combination of dust size and amorphous carbon optical constants was able to match how red the spectrum is -- but fit the 2022 February outburst spectrum about equally as well as the Mie scattering model. Over the wavelength range of $0.8-1.6\mu{m}$ the two models perform similarly, the Hapke model performs better in the (noisy) $1.60-1.77\mu{m}$ range, and the Mie model performs better at wavelengths longer than $2.0\mu{m}$. In other words, neither model is perfect, but the Mie model fails somewhat less. One simple option is that the material in the 2022 February outburst reflected light closer to how large grains alone would reflect light compared to the material in the 2017 August outburst which reflected light more like how fine particles would. The Mie scattering models also suggest that the differences between these two spectra come from having very different grain size distributions. The 2017 outburst was modeled with a steep size distribution in which the smallest grains completely dominate the reflective cross section of the coma, while the 2022 outburst was modeled with a much shallower size distribution where small grains are still important, just less so.

What are the origin of these differences, and do they provide physical insight into how outbursts proceed on SW1? One option is that these two outbursts are simply different for intrinsic reasons. Perhaps they were initiated on parts of SW1's surface with very different grain size distributions (or different grain composition which mimics grain size distribution differences) or maybe smaller outbursts at SW1 are powered or caused by different processes than large ones. (We note that both of these outbursts would be considered significant at any less outburst-prone comet.) While the former scenario is challenging to assess the likelihood of (we know very little about SW1's surface past size and average thermal emission properties, \citealt{2021PSJ.....2..126S}), the latter has at least some arguments against it. Would a larger and more energetic outburst be able to loft more larger grains, and thus affect a shallower size distribution? Why would a weaker outburst lift grains with a larger average size? If the two outbursts were produced by completely different processes -- say, amorphous ice crystallization \citep{2022PSJ.....3..251L} and the collapse of surface topography like cliffs or overhangs (see, e.g., \citealt{2016Icar..272...60S, 2017NatAs...1E..92P, 2021AJ....162....4N}) -- then differences in how the expanding dust is accelerated might be expected. That said, the models still indicate that one outburst observation requires significantly more fine grains than the other, and this might be a clue to which process powered each outburst.

A scenario which could explain most of these differences is related to \textit{when} we observed SW1 relative to the outburst. In the case of the 2017 August outburst, our spectra were obtained well before the coma reached peak brightness -- possibly before the dust released during the impulsive event had spread out enough to become optically thin. The 2022 February outburst is the opposite -- the spectra were obtained after the coma had become thin enough that all of the grains were illuminated and the coma had thus reached peak brightness. Thus, even if the two outbursts ejected dust with the same size distribution and composition, our 2017 August spectrum would have only `seen' the outermost dust ejected in the outburst whereas the 2022 February spectrum would have seen a more representative sampling of the bulk outburst ejecta on top of the pre-existing quiescent coma. This scenario requires the outermost regions of the expanding outburst to have a different size distribution than the outburst as a whole. This is a natural consequence of any outburst in which the expanding dust is accelerated by outflowing gas (the acceleration of a given grain should scale as the inverse of the grain's radius), such that smaller grains are accelerated more significantly. The fact that our 2017 August outburst spectrum is best fit by such a steep particle distribution is then a function of when we observed it, not necessarily that it ejected a higher fraction of small grains than our 2022 February outburst. In essence, we argue that we can explain the differences in our two outburst spectra as an optical depth effect without requiring any intrinsic differences in the kinds of materials ejected during the outbursts.

A confounding factor to this interpretation, though one that would primarily affect observations taken early in an outburst, is that if the nucleus contributes significant light to the quiescent coma (it likely does, see next sub-section), the early stages of an outburst are occurring `on top' of the nucleus and the quiescent coma. Early in the outburst, the new dust is both reflecting sunlight and obscuring dust which was previously illuminated. A one-component model for what kind of surface is reflecting light is just not complex enough to capture that kind of nuance -- but the large difference between our strongest outburst (Aug. 2017) and the other data suggests this is likley not significant for our four datasets. We discuss this kind of complication further in the next sub-section.

\citet{2016Icar..272..327M} found for one of the outbursts in their study that at wavelengths beyond $\sim0.7\mu{m}$, the reflectance of SW1's coma became more neutral as determined from broadband photometry. If our two outburst spectra are indeed snapshots of the same process seen at different times, then we would draw the same conclusion from our data -- the spectra became more neutral over time. A more targeted ToO campaign to capture a large outburst of SW1 several times before and after it reached peak brightness seems like the best way to test the validity of this scenario. If the spectrum begins deeply red and concave, like our 2017 August outburst spectrum, and proceeds towards a more neutral and linear spectrum, like our 2022 February outburst spectrum, this would be extremely good evidence that our hypothesis is on the right track. In the near-infrared, this might be possible with just (Y-)J-H-K imaging -- and thus less time intensive than the observations probed here. Broadband imaging would be less sensitive to narrow features like the absorption bands of water ice, so there would be an implicit fitting degeneracy between changes in the dust size distribution and water ice abundance/properties that would be harder to disentangle. Campaigns at other wavelengths, like in the blue/near-UV, might be effective as well -- but $CO^+$ contamination at blue wavelengths might be challenging to account for with broadband photometry alone.

If the spectral variation we see is an optical-depth/timing effect as we have proposed, a future line of research would be to attempt to invert the properties of the gas flow that accelerated and size-sorted the grains from the rate at which the spectrum evolved over time. Do weaker outbursts sort their grains less clearly than more energetic ones? If possible, this kind of analysis might allow for a constraint on gas production rates and outflow speeds that is only available from the ground from millimeter wavelengths otherwise \citep{2020AJ....159..136W, morvan_2022}.

\subsection{Quiescent activity}
The two quiescent spectra we obtained for this study appear rather similar and can be fit acceptably by either large-grained Hapke models or fine-grained Mie scattering models with very similar parameters. While the Hapke models were rather unconstrained in grain size, neither needed significant ice to achieve a good fit. The Mie scattering models converged on size distribution power laws between $\alpha=4.1-5.2$, which are similar to what has been found for other comets undergoing stable ongoing activity using these kinds of models \citep{2018ApJ...862L..16P, 2023PSJ.....4...85K}. In some sense, SW1's quiescent activity looks like stable activity at other comets.

While SW1 does not appear to ever fully shut off and become inactive like some other outburst-prone Centaurs \citep{2019AJ....158..255K}, the absolute magnitude of its nucleus (assuming the size and albedo of \citealt{2021PSJ.....2..126S}) is $H_V\sim10$ -- not much fainter than the quiescent magnitudes reported by ATLAS or those reported to the BAA and shown in Figure \ref{fig:log}. Those photometric observations were taken with larger apertures than one would use for a point source, so the contribution from the nucleus in smaller apertures would be somewhat higher. One possibility for why the two quiescent spectra can apparently be equally well described as being composed of large grains or a conventional mix of grain sizes might just be that the nuclear contamination is non-negligible. If a substantial fraction of the light from SW1 at visual wavelengths (where absolute magnitudes are defined) is reflected light from the nucleus, the nucleus should be increasingly important at longer wavelengths as smaller grains lose their relevance. While a half-magnitude difference between the total brightness of the comet and nucleus alone might imply that some two-thirds of the light one receives is from the nucleus, that would actually be an upper limit as dust grains `in front' of the nucleus would obscure the nucleus at an unknown level. This adds to our inability to interpret the long-wavelength reflectivity of our low-activity spectra. Given that the comet is angularly small in the near-infrared as mentioned in Section \ref{sec:obs}, one cannot dither the telescope around the coma as easily as at shorter wavelengths to look for spectral variations that might be due to this in the very innermost coma excepting better seeing than we achieved here.

As noted throughout Section \ref{sec:models}, there is a very real chance that our models are just not sophisticated enough to capture all of the reflective behavior seen. Some of this might be the nucleus, but a more thorough ruling-out or accounting-for of more complex grain size distributions (with multiple power laws or populations of grains) would be needed before nuclear contamination could be confidently assessed. That said, if one assumes that the grains in the vicinity of SW1's nucleus far from any outburst are those larger grains least effected by radiation pressure, one could make a reasonable assumption that a \citet{2012tres.book.....H} model would be equally applicable for the nucleus and the remnant dust. Future modelers could use an approach in which the reflected light from the nucleus and the large grained coma are modeled separately and then combined and weighted by the visual brightness of the comet and each of their respective visual albedos. One would have to assume or fit for the amount of `shadowing' described above, but it might result in more reliable constraints on the dust sizes in the coma than were possible with our approach. In any case, if the nucleus reflects light like larger grains and contributes any significant light to our quiescent spectra, it is likely making the size distribution we infer look slightly flatter than it really is.

\subsection{Where is the water coming from?}
The two quiescent spectra were both best fit with about a half-percent by area of ice in the coma, though one can tell by even a brief inspection of the spectra themselves that this is hardly a very strong detection. Our larger outburst shows less evidence for ice, while our later observation of a smaller outburst might show `stronger' evidence for water ice if the dust grains were large. (Hapke models preferred an areal ice fraction of about $1.4\%$, while Mie models preferred no ice.) However, \textit{Herschel} observations reported and analyzed in \citet{2022A&A...664A..95B} suggest that the water production seen at SW1 ($(4.1-5.9)\times10^{27}$ molecules per second, with this range including their observations and those of \citealt{2012ApJ...752...15O}) is primarily coming from its icy grain population, estimating a total area which would need to be sublimating of some $\sim2000km^2$ -- about 440\% the surface area of the nucleus. In our quiescent spectra, the total brightness of SW1 was about $\sim50\%$ higher than reflected light from the nucleus alone would be, which when combined with a small ice fraction in any of our spectra makes it challenging to imagine that there were enough icy grains in our aperture to be able to explain those previous observations. Even for our 2022 outburst observations, where the brightness of the coma was about $750\%$ of the bare nucleus, this would require an ice fraction at least an order of magnitude higher than our most optimistic ice fraction estimates. Two possibilities could reconcile their requirement for icy grains and our non-detections of them. The first is that the ice is mixed with opaque materials like carbon at such a level that its spectral signatures could be masked. We did attempt mixed dust-and-ice models (as mentioned earlier in Section 3.1) as with limited success -- the fits were of somewhat worse quality, but a $2-4\%$ volume fraction of ice could be accommodated in the quiescent spectra while still not displaying any obvious ice absorption features. The second possibility would be that the ice is in much larger grains that are not as optically important at the wavelengths of interest. The real answer likely lies in a combination of both scenarios. If the ice was primarily stored in both larger and mixed grains, this might be functionally invisible to our spectra and thus be a way to reconcile their observations with our own. Interestingly, the icy grains around P/2019 LD2 (ATLAS), another object that only recently left an SW1-like orbit and thermal environment, shows evidence for just such large mixed grains \citep{2021PSJ.....2...48K}, likely in part because its lack of smaller grains made the larger grains easier to detect.

\section{Summary} \label{sec:summary}
29P/Schwassmann-Wachmann 1 (``SW1") is best known for its outbursts. Despite a nearly circular orbit just beyond  Jupiter, and thus limited variation in limited solar insolation, the object undergoes several large multiple-magnitude outbursts every year and has done so since discovery in 1927. This paper reports and models four epochs of low-resolution near-infrared spectra taken between 2017 and 2022 at the NASA Infrared Telescope Facility and the Lowell Discovery Telescope. This dataset includes a large composite outburst in 2017, a `quiescent' or low-activity period in 2020, another quiescent observation in February 2022 followed days later by an observation of the comet after a moderate outburst. Through comparing these spectra and our analyses of them, we come to the two following primary conclusions:

\begin{itemize}
    \item The two outburst spectra represent the reddest and most spectrally-neutral end members of the dataset, while the two quiescent spectra appear consistent in average spectral slope and overall spectral behavior. The quiescent spectra are both moderately red in spectral slope, and we infer that SW1's spectral behavior during quiescent activity might be relatively stable -- and thus that whatever ongoing process powers the quiescent activity at SW1 is similarly stable. Our spectral models infer a relatively typical size distribution of dust in the quiescent periods similar to other comets in stable activity. In other words, quiescent activity at SW1 looks like traditional activity at other comets -- SW1 looks like a `typical' comet when it is not outbursting.
    \item The difference in spectra between the two outbursts might be that the two events released fundamentally different materials, or it could be an optical depth effect. Given the apparent short duration of these outbursts, there must be some period as the material is expanding outwards where the coma is briefly optically thick and only brightening as the material thins and more dust is uncovered. The two outburst spectra were obtained before and after the peak in brightness of their respective outbursts, so we might have seen the `outermost' material in the spectrum obtained before the peak and the `whole' coma in the one seen afterwards. The differences between the two spectra, as interpreted through our models, would imply a size-sorting of the grains, such that finer grains end up on the outside, and thus a time when the gas and dust are coupled in these outbursts such that the finer grains are moving faster. Given that the duration of these outbursts must be very short compared to the rotation period of the nucleus (see, e.g., \citealt{schambeau_2017}), this places a quantitative constraint on the minimum time in which the gas and dust are fully coupled in these explosive events.
\end{itemize}

\begin{acknowledgments}
TK was supported in part by Lowell Observatory internal funds, including but not limited to the Lowell Observatory Marcus Comet Research Fund.

\end{acknowledgments}

%

\vspace{5mm}
\facilities{IRTF(SpeX), LDT(NIHTS)}


\software{}



\appendix

\section{Appendix information}



\begin{thebibliography}{}
\expandafter\ifx\csname natexlab\endcsname\relax\def\natexlab#1{#1}\fi
\providecommand{\url}[1]{\href{#1}{#1}}
\providecommand{\dodoi}[1]{doi:~\href{http://doi.org/#1}{\nolinkurl{#1}}}
\providecommand{\doeprint}[1]{\href{http://ascl.net/#1}{\nolinkurl{http://ascl.net/#1}}}
\providecommand{\doarXiv}[1]{\href{https://arxiv.org/abs/#1}{\nolinkurl{https://arxiv.org/abs/#1}}}

\bibitem[{{Bauer} {et~al.}(2008){Bauer}, {Choi}, {Weissman}, {Stansberry},
  {Fern{\'a}ndez}, {Roe}, {Buratti}, \& {Sung}}]{2008PASP..120..393B}
{Bauer}, J.~M., {Choi}, Y.-J., {Weissman}, P.~R., {et~al.} 2008, \pasp, 120,
  393, \dodoi{10.1086/587552}

\bibitem[{{Bauer} {et~al.}(2017){Bauer}, {Grav}, {Fern{\'a}ndez}, {Mainzer},
  {Kramer}, {Masiero}, {Spahr}, {Nugent}, {Stevenson}, {Meech}, {Cutri},
  {Lisse}, {Walker}, {Dailey}, {Rosser}, {Krings}, {Ruecker}, {Wright}, \&
  {NEOWISE Team}}]{bauer_2017}
{Bauer}, J.~M., {Grav}, T., {Fern{\'a}ndez}, Y.~R., {et~al.} 2017, \aj, 154,
  53, \dodoi{10.3847/1538-3881/aa72df}

\bibitem[{{Bockel{\'e}e-Morvan}
  {et~al.}(2022{\natexlab{a}}){Bockel{\'e}e-Morvan}, {Biver}, {Schambeau},
  {Crovisier}, {Opitom}, {de Val Borro}, {Lellouch}, {Hartogh},
  {Vandenbussche}, {Jehin}, {Kidger}, {K{\"u}ppers}, {Lis}, {Moreno},
  {Szutowicz}, \& {Zakharov}}]{morvan_2022}
{Bockel{\'e}e-Morvan}, D., {Biver}, N., {Schambeau}, C.~A., {et~al.}
  2022{\natexlab{a}}, \aap, 664, A95, \dodoi{10.1051/0004-6361/202243241}

\bibitem[{{Bockel{\'e}e-Morvan}
  {et~al.}(2022{\natexlab{b}}){Bockel{\'e}e-Morvan}, {Biver}, {Schambeau},
  {Crovisier}, {Opitom}, {de Val-Borro}, {Lellouch}, {Hartogh},
  {Vandenbussche}, {Jehin}, {Kidger}, {K{\"u}ppers}, {Lis}, {Moreno},
  {Szutowicz}, \& {Zakharov}}]{2022A&A...664A..95B}
---. 2022{\natexlab{b}}, \aap, 664, A95, \dodoi{10.1051/0004-6361/202243241}

\bibitem[{Bruggeman(1935)}]{bruggeman1935berechnung}
Bruggeman, V.~D. 1935, Annalen der physik, 416, 636

\bibitem[{{Cochran} {et~al.}(1980){Cochran}, {Barker}, \&
  {Cochran}}]{1980AJ.....85..474C}
{Cochran}, A., {Barker}, E.~S., \& {Cochran}, W. 1980, \aj, 85, 474,
  \dodoi{10.1086/112699}

\bibitem[{{Cochran} \& {Cochran}(1991)}]{1991Icar...90..172C}
{Cochran}, A.~L., \& {Cochran}, W.~D. 1991, \icarus, 90, 172,
  \dodoi{10.1016/0019-1035(91)90077-7}

\bibitem[{{Cochran} {et~al.}(1982){Cochran}, {Cochran}, \&
  {Barker}}]{1982ApJ...254..816C}
{Cochran}, A.~L., {Cochran}, W.~D., \& {Barker}, E.~S. 1982, \apj, 254, 816,
  \dodoi{10.1086/159793}

\bibitem[{{Cushing} {et~al.}(2004){Cushing}, {Vacca}, \&
  {Rayner}}]{2004PASP..116..362C}
{Cushing}, M.~C., {Vacca}, W.~D., \& {Rayner}, J.~T. 2004, \pasp, 116, 362,
  \dodoi{10.1086/382907}

\bibitem[{{Edoh}(1983)}]{1983PhDT........18E}
{Edoh}, O. 1983, PhD thesis, University of Arizona

\bibitem[{{Fern{\'a}ndez} {et~al.}(2013){Fern{\'a}ndez}, {Kelley}, {Lamy},
  {Toth}, {Groussin}, {Lisse}, {A'Hearn}, {Bauer}, {Campins}, {Fitzsimmons},
  {Licand ro}, {Lowry}, {Meech}, {Pittichov{\'a}}, {Reach}, {Snodgrass}, \&
  {Weaver}}]{fernandez_2013}
{Fern{\'a}ndez}, Y.~R., {Kelley}, M.~S., {Lamy}, P.~L., {et~al.} 2013, \icarus,
  226, 1138, \dodoi{10.1016/j.icarus.2013.07.021}

\bibitem[{{Firgard} \& {Kareta}(2025)}]{firgard_and_kareta}
{Firgard}, M., \& {Kareta}, T. 2025, In review at /psj

\bibitem[{{Foreman-Mackey} {et~al.}(2013){Foreman-Mackey}, {Hogg}, {Lang}, \&
  {Goodman}}]{2013PASP..125..306F}
{Foreman-Mackey}, D., {Hogg}, D.~W., {Lang}, D., \& {Goodman}, J. 2013, \pasp,
  125, 306, \dodoi{10.1086/670067}

\bibitem[{{Guilbert-Lepoutre} {et~al.}(2023){Guilbert-Lepoutre}, {Gkotsinas},
  {Raymond}, \& {Nesvorny}}]{2023ApJ...942...92G}
{Guilbert-Lepoutre}, A., {Gkotsinas}, A., {Raymond}, S.~N., \& {Nesvorny}, D.
  2023, \apj, 942, 92, \dodoi{10.3847/1538-4357/acaa3a}

\bibitem[{{Gustafsson} {et~al.}(2021){Gustafsson}, {Moskovitz}, {Cushing},
  {Bida}, {Dunham}, \& {Roe}}]{2021PASP..133c5001G}
{Gustafsson}, A., {Moskovitz}, N., {Cushing}, M.~C., {et~al.} 2021, \pasp, 133,
  035001, \dodoi{10.1088/1538-3873/abe2f4}

\bibitem[{{Hapke}(2012)}]{2012tres.book.....H}
{Hapke}, B. 2012, {Theory of Reflectance and Emittance Spectroscopy},
  \dodoi{10.1017/CBO9781139025683}

\bibitem[{{Heinze} {et~al.}(2018){Heinze}, {Tonry}, {Denneau}, {Flewelling},
  {Stalder}, {Rest}, {Smith}, {Smartt}, \& {Weiland}}]{2018AJ....156..241H}
{Heinze}, A.~N., {Tonry}, J.~L., {Denneau}, L., {et~al.} 2018, \aj, 156, 241,
  \dodoi{10.3847/1538-3881/aae47f10.48550/arXiv.1804.02132}

\bibitem[{{Ivanova} {et~al.}(2019){Ivanova}, {Agapitov}, {Odstrcil}, {Korsun},
  {Afanasiev}, \& {Rosenbush}}]{2019MNRAS.486.5614I}
{Ivanova}, O., {Agapitov}, O., {Odstrcil}, D., {et~al.} 2019, \mnras, 486,
  5614, \dodoi{10.1093/mnras/stz1200}

\bibitem[{{Jewitt}(1990)}]{1990ApJ...351..277J}
{Jewitt}, D. 1990, \apj, 351, 277, \dodoi{10.1086/168463}

\bibitem[{{Kareta} {et~al.}(2023){Kareta}, {Noonan}, {Harris}, \&
  {Springmann}}]{2023PSJ.....4...85K}
{Kareta}, T., {Noonan}, J.~W., {Harris}, W.~M., \& {Springmann}, A. 2023, \psj,
  4, 85, \dodoi{10.3847/PSJ/accc28}

\bibitem[{{Kareta} {et~al.}(2019){Kareta}, {Sharkey}, {Noonan}, {Volk},
  {Reddy}, {Harris}, \& {Miles}}]{2019AJ....158..255K}
{Kareta}, T., {Sharkey}, B., {Noonan}, J., {et~al.} 2019, \aj, 158, 255,
  \dodoi{10.3847/1538-3881/ab505f}

\bibitem[{{Kareta} {et~al.}(2021){Kareta}, {Woodney}, {Schambeau}, {Fernandez},
  {Harrington Pinto}, {Wierzchos}, {Womack}, {Bus}, {Steckloff}, {Sarid},
  {Volk}, {Harris}, \& {Reddy}}]{2021PSJ.....2...48K}
{Kareta}, T., {Woodney}, L.~M., {Schambeau}, C., {et~al.} 2021, \psj, 2, 48,
  \dodoi{10.3847/PSJ/abe23d}

\bibitem[{{Larson}(1980)}]{1980ApJ...238L..47L}
{Larson}, S.~M. 1980, \apjl, 238, L47, \dodoi{10.1086/183255}

\bibitem[{{Lisse} {et~al.}(2022){Lisse}, {Steckloff}, {Prialnik}, {Womack},
  {Harrington Pinto}, {Sarid}, {Fernandez}, {Schambeau}, {Kareta},
  {Samarasinha}, {Harris}, {Volk}, {Woodney}, {Cruikshank}, \&
  {Sandford}}]{2022PSJ.....3..251L}
{Lisse}, C.~M., {Steckloff}, J.~K., {Prialnik}, D., {et~al.} 2022, \psj, 3,
  251, \dodoi{10.3847/PSJ/ac9468}

\bibitem[{{Marsset} {et~al.}(2020){Marsset}, {DeMeo}, {Binzel}, {Bus},
  {Burbine}, {Burt}, {Moskovitz}, {Polishook}, {Rivkin}, {Slivan}, \&
  {Thomas}}]{2020ApJS..247...73M}
{Marsset}, M., {DeMeo}, F.~E., {Binzel}, R.~P., {et~al.} 2020, \apjs, 247, 73,
  \dodoi{10.3847/1538-4365/ab7b5f}

\bibitem[{{Mastrapa} {et~al.}(2008){Mastrapa}, {Bernstein}, {Sandford},
  {Roush}, {Cruikshank}, \& {Dalle Ore}}]{2008Icar..197..307M}
{Mastrapa}, R.~M., {Bernstein}, M.~P., {Sandford}, S.~A., {et~al.} 2008,
  \icarus, 197, 307, \dodoi{10.1016/j.icarus.2008.04.008}

\bibitem[{{Miles} \& {BAA Comet Section 29P Observing
  Campaign}(2022)}]{2022DPS....5441403M}
{Miles}, R., \& {BAA Comet Section 29P Observing Campaign}. 2022, in
  AAS/Division for Planetary Sciences Meeting Abstracts, Vol.~54, AAS/Division
  for Planetary Sciences Meeting Abstracts, 414.03

\bibitem[{{Miles} {et~al.}(2016){Miles}, {Faillace}, {Mottola}, {Raab},
  {Roche}, {Soulier}, \& {Watkins}}]{2016Icar..272..327M}
{Miles}, R., {Faillace}, G.~A., {Mottola}, S., {et~al.} 2016, \icarus, 272,
  327, \dodoi{10.1016/j.icarus.2015.11.019}

\bibitem[{{Montalto} {et~al.}(2008){Montalto}, {Riffeser}, {Hopp}, {Wilke}, \&
  {Carraro}}]{2008A&A...479L..45M}
{Montalto}, M., {Riffeser}, A., {Hopp}, U., {Wilke}, S., \& {Carraro}, G. 2008,
  \aap, 479, L45, \dodoi{10.1051/0004-6361:20079130}

\bibitem[{{Noonan} {et~al.}(2021){Noonan}, {Rinaldi}, {Feldman}, {Stern},
  {Parker}, {Keeney}, {Bockel{\'e}e-Morvan}, {Vervack}, {Steffl}, {Knight},
  {Schindhelm}, {Feaga}, {Pineau}, {Medina}, {Weaver}, {Bertaux}, \&
  {A'Hearn}}]{2021AJ....162....4N}
{Noonan}, J.~W., {Rinaldi}, G., {Feldman}, P.~D., {et~al.} 2021, \aj, 162, 4,
  \dodoi{10.3847/1538-3881/abf8b4}

\bibitem[{{Ootsubo} {et~al.}(2012){Ootsubo}, {Kawakita}, {Hamada}, {Kobayashi},
  {Yamaguchi}, {Usui}, {Nakagawa}, {Ueno}, {Ishiguro}, {Sekiguchi}, {Watanabe},
  {Sakon}, {Shimonishi}, \& {Onaka}}]{2012ApJ...752...15O}
{Ootsubo}, T., {Kawakita}, H., {Hamada}, S., {et~al.} 2012, \apj, 752, 15,
  \dodoi{10.1088/0004-637X/752/1/15}

\bibitem[{{Pajola} {et~al.}(2017){Pajola}, {H{\"o}fner}, {Vincent}, {Oklay},
  {Scholten}, {Preusker}, {Mottola}, {Naletto}, {Fornasier}, {Lowry}, {Feller},
  {Hasselmann}, {G{\"u}ttler}, {Tubiana}, {Sierks}, {Barbieri}, {Lamy},
  {Rodrigo}, {Koschny}, {Rickman}, {Keller}, {Agarwal}, {A'Hearn}, {Barucci},
  {Bertaux}, {Bertini}, {Besse}, {Boudreault}, {Cremonese}, {da Deppo},
  {Davidsson}, {Debei}, {de Cecco}, {Deller}, {Deshapriya}, {El-Maarry},
  {Ferrari}, {Ferri}, {Fulle}, {Groussin}, {Gutierrez}, {Hofmann}, {Hviid},
  {Ip}, {Jorda}, {Knollenberg}, {Kovacs}, {Kramm}, {K{\"u}hrt}, {K{\"u}ppers},
  {Lara}, {Lin}, {Lazzarin}, {Lucchetti}, {Lopez Moreno}, {Marzari},
  {Massironi}, {Michalik}, {Penasa}, {Pommerol}, {Simioni}, {Thomas}, {Toth},
  \& {Baratti}}]{2017NatAs...1E..92P}
{Pajola}, M., {H{\"o}fner}, S., {Vincent}, J.~B., {et~al.} 2017, Nature
  Astronomy, 1, 0092, \dodoi{10.1038/s41550-017-0092}

\bibitem[{{Protopapa} {et~al.}(2018){Protopapa}, {Kelley}, {Yang}, {Bauer},
  {Kolokolova}, {Woodward}, {Keane}, \& {Sunshine}}]{2018ApJ...862L..16P}
{Protopapa}, S., {Kelley}, M. S.~P., {Yang}, B., {et~al.} 2018, \apjl, 862,
  L16, \dodoi{10.3847/2041-8213/aad33b}

\bibitem[{{Rayner} {et~al.}(2003){Rayner}, {Toomey}, {Onaka}, {Denault},
  {Stahlberger}, {Vacca}, {Cushing}, \& {Wang}}]{2003PASP..115..362R}
{Rayner}, J.~T., {Toomey}, D.~W., {Onaka}, P.~M., {et~al.} 2003, \pasp, 115,
  362, \dodoi{10.1086/367745}

\bibitem[{{Roemer}(1958)}]{1958PASP...70..272R}
{Roemer}, E. 1958, \pasp, 70, 272, \dodoi{10.1086/127223}

\bibitem[{{Rouleau} \& {Martin}(1991)}]{1991ApJ...377..526R}
{Rouleau}, F., \& {Martin}, P.~G. 1991, \apj, 377, 526, \dodoi{10.1086/170382}

\bibitem[{{Sarid} {et~al.}(2019){Sarid}, {Volk}, {Steckloff}, {Harris},
  {Womack}, \& {Woodney}}]{2019ApJ...883L..25S}
{Sarid}, G., {Volk}, K., {Steckloff}, J.~K., {et~al.} 2019, \apjl, 883, L25,
  \dodoi{10.3847/2041-8213/ab3fb3}

\bibitem[{{Schambeau} {et~al.}(2017){Schambeau}, {Fern{\'a}ndez},
  {Samarasinha}, {Mueller}, \& {Woodney}}]{schambeau_2017}
{Schambeau}, C.~A., {Fern{\'a}ndez}, Y.~R., {Samarasinha}, N.~H., {Mueller}, B.
  E.~A., \& {Woodney}, L.~M. 2017, \icarus, 284, 359,
  \dodoi{10.1016/j.icarus.2016.11.026}

\bibitem[{{Schambeau} {et~al.}(2021){Schambeau}, {Fern{\'a}ndez},
  {Samarasinha}, {Womack}, {Bockel{\'e}e-Morvan}, {Lisse}, \&
  {Woodney}}]{2021PSJ.....2..126S}
{Schambeau}, C.~A., {Fern{\'a}ndez}, Y.~R., {Samarasinha}, N.~H., {et~al.}
  2021, \psj, 2, 126, \dodoi{10.3847/PSJ/abfe6f}

\bibitem[{{Schambeau} {et~al.}(2019){Schambeau}, {Fern{\'a}ndez},
  {Samarasinha}, {Woodney}, \& {Kundu}}]{schambeau_2019}
{Schambeau}, C.~A., {Fern{\'a}ndez}, Y.~R., {Samarasinha}, N.~H., {Woodney},
  L.~M., \& {Kundu}, A. 2019, \aj, 158, 259, \dodoi{10.3847/1538-3881/ab53e2}

\bibitem[{{Senay} \& {Jewitt}(1994)}]{1994Natur.371..229S}
{Senay}, M.~C., \& {Jewitt}, D. 1994, \nat, 371, 229, \dodoi{10.1038/371229a0}

\bibitem[{{Shingles} {et~al.}(2021){Shingles}, {Smith}, {Young}, {Smartt},
  {Tonry}, {Denneau}, {Heinze}, {Weiland}, {Flewelling}, {Stalder},
  {Clocchiatti}, {F{\"o}rster}, {Pignata}, {Rest}, {Anderson}, {Stubbs}, \&
  {Erasmus}}]{2021TNSAN...7....1S}
{Shingles}, L., {Smith}, K.~W., {Young}, D.~R., {et~al.} 2021, Transient Name
  Server AstroNote, 7, 1

\bibitem[{{Smith} {et~al.}(2020){Smith}, {Smartt}, {Young}, {Tonry}, {Denneau},
  {Flewelling}, {Heinze}, {Weiland}, {Stalder}, {Rest}, {Stubbs}, {Anderson},
  {Chen}, {Clark}, {Do}, {F{\"o}rster}, {Fulton}, {Gillanders}, {McBrien},
  {O'Neill}, {Srivastav}, \& {Wright}}]{2020PASP..132h5002S}
{Smith}, K.~W., {Smartt}, S.~J., {Young}, D.~R., {et~al.} 2020, \pasp, 132,
  085002, \dodoi{10.1088/1538-3873/ab936e10.48550/arXiv.2003.09052}

\bibitem[{{Steckloff} {et~al.}(2016){Steckloff}, {Graves}, {Hirabayashi},
  {Melosh}, \& {Richardson}}]{2016Icar..272...60S}
{Steckloff}, J.~K., {Graves}, K., {Hirabayashi}, M., {Melosh}, H.~J., \&
  {Richardson}, J.~E. 2016, \icarus, 272, 60,
  \dodoi{10.1016/j.icarus.2016.02.026}

\bibitem[{{Steckloff} {et~al.}(2020){Steckloff}, {Sarid}, {Volk}, {Kareta},
  {Womack}, {Harris}, {Woodney}, \& {Schambeau}}]{2020ApJ...904L..20S}
{Steckloff}, J.~K., {Sarid}, G., {Volk}, K., {et~al.} 2020, \apjl, 904, L20,
  \dodoi{10.3847/2041-8213/abc888}

\bibitem[{{Sunshine} {et~al.}(2007){Sunshine}, {Groussin}, {Schultz},
  {A'Hearn}, {Feaga}, {Farnham}, \& {Klaasen}}]{2007Icar..190..284S}
{Sunshine}, J.~M., {Groussin}, O., {Schultz}, P.~H., {et~al.} 2007, \icarus,
  190, 284, \dodoi{10.1016/j.icarus.2007.04.024}

\bibitem[{{Tonry} {et~al.}(2018){Tonry}, {Denneau}, {Heinze}, {Stalder},
  {Smith}, {Smartt}, {Stubbs}, {Weiland}, \& {Rest}}]{2018PASP..130f4505T}
{Tonry}, J.~L., {Denneau}, L., {Heinze}, A.~N., {et~al.} 2018, \pasp, 130,
  064505, \dodoi{10.1088/1538-3873/aabadf10.48550/arXiv.1802.00879}

\bibitem[{{Trigo-Rodr{\'\i}guez} {et~al.}(2010){Trigo-Rodr{\'\i}guez},
  {Garc{\'\i}a-Hern{\'a}ndez}, {S{\'a}nchez}, {Lacruz}, {Davidsson},
  {Rodr{\'\i}guez}, {Pastor}, \& {de Los Reyes}}]{2010MNRAS.409.1682T}
{Trigo-Rodr{\'\i}guez}, J.~M., {Garc{\'\i}a-Hern{\'a}ndez}, D.~A.,
  {S{\'a}nchez}, A., {et~al.} 2010, \mnras, 409, 1682,
  \dodoi{10.1111/j.1365-2966.2010.17425.x}

\bibitem[{{Trigo-Rodr{\'\i}guez} {et~al.}(2008){Trigo-Rodr{\'\i}guez},
  {Garc{\'\i}a-Melendo}, {Davidsson}, {S{\'a}nchez}, {Rodr{\'\i}guez},
  {Lacruz}, {de Los Reyes}, \& {Pastor}}]{2008A&A...485..599T}
{Trigo-Rodr{\'\i}guez}, J.~M., {Garc{\'\i}a-Melendo}, E., {Davidsson},
  B.~J.~R., {et~al.} 2008, \aap, 485, 599, \dodoi{10.1051/0004-6361:20078666}

\bibitem[{{Whipple}(1980)}]{1980AJ.....85..305W}
{Whipple}, F.~L. 1980, \aj, 85, 305, \dodoi{10.1086/112676}

\bibitem[{{Wierzchos} \& {Womack}(2020)}]{2020AJ....159..136W}
{Wierzchos}, K., \& {Womack}, M. 2020, \aj, 159, 136,
  \dodoi{10.3847/1538-3881/ab6e68}

\end{thebibliography}

\bibliographystyle{aasjournal}



\end{document}